\documentclass{article}
\usepackage{spconf,amsmath,amssymb,graphicx,booktabs}
\usepackage{hyperref}
\usepackage{flafter}

\graphicspath{{./}}
\usepackage{pifont}
\usepackage{cite}
\usepackage{multirow}
\usepackage{stfloats}
\usepackage{capt-of}
\usepackage{xspace}
\newcommand{\cmark}{\ding{51}}
\newcommand{\xmark}{\ding{55}}
\title{Exemplar-Free Analytic Learning for Multi-Label Audio Class-Incremental Learning}

\name{%
Siyuan Luo, Yang Xiao, Ting Dang
}
\address{The University of Melbourne, Australia}

\newif\ifshowcomments
\showcommentstrue
\ifdefined\HideReviewComments\showcommentsfalse\fi
\usepackage{soul, xcolor}
\soulregister\cite7
\soulregister\ref7

\newcommand{\method}{ALMA\xspace}

\begin{document}
\ninept
\setlength{\parskip}{0pt}
\setlength{\abovedisplayskip}{4pt plus 1pt minus 1pt}
\setlength{\belowdisplayskip}{4pt plus 1pt minus 1pt}
\setlength{\abovedisplayshortskip}{2pt plus 1pt}
\setlength{\belowdisplayshortskip}{3pt plus 1pt minus 1pt}
\setlength{\textfloatsep}{7pt plus 2pt minus 2pt}
\setlength{\floatsep}{6pt plus 2pt minus 2pt}
\setlength{\abovecaptionskip}{3pt}
\setlength{\belowcaptionskip}{0pt}
\maketitle

\begin{abstract}
Audio classification is inherently a multi-label task, as real-world
acoustic environments contain multiple simultaneous sound events. When
new sound classes emerge, models must incorporate them
without forgetting previously learned ones: a challenge known as class-incremental learning. Existing methods rely on storing past data and iterative gradient
updates, which struggle under incomplete multi-label supervision because
only the newly introduced classes are annotated at each phase, leaving old-class labels unavailable. We investigate exemplar-free analytic continual learning as a principled alternative, in which a linear classifier is updated in closed form without storing historical recordings or performing incremental back-propagation, and previously learned weights remain intact by construction. Building on analytic learning, we further propose \method, which addresses incomplete supervision and class imbalance through continuous old-class score estimates and
frequency-based sample weighting. Experiments on a 50-class AudioSet-R benchmark across three incremental setups show that the analytic
learner substantially outperforms gradient-based methods, and previously learned classes retain nearly unchanged detection performance as new
classes are added. This study shows that \method is a simple yet effective solution to multi-label audio class-incremental learning. 

\end{abstract}


\section{Introduction}
Sound recognition systems in smart homes, factories and ecological monitoring may face new sound classes after deployment~\cite{dekkers2017sins,purohit2019mimii,kahl2021birdnet, xiao2022continual}, such as breaking glass, machine
faults or newly observed wildlife calls.
While single-label audio classification has advanced rapidly~\cite{mohmmad2024sedreview}, it ignores the inherent
nature of sound co-occurrence in real-world environments~\cite{gemmeke2017audioset}.
For instance, in a kitchen environment, sounds such as running water, sizzling, and speech naturally overlap simultaneously, a pattern that single-label models are inherently unable to capture. Learning sounds in isolation inevitably fails to represent these co-occurrence relationships. Multi-label audio classification emerges as a natural solution by predicting multiple sound events simultaneously~\cite{joshi2018timeaggregation}; 
however, the space of possible co-occurrence patterns is effectively unbounded in real world, necessitating that models continuously incorporate newly arriving sound classes without forgetting previously learned ones: a challenge referred to as multi-label audio class-incremental learning (CIL)~\cite{mulimani2024mlaudiocil}.


Multi-label audio continual learning is especially challenging because
each incremental phase provides labels only for its newly introduced
class group~\cite{mulimani2024mlaudiocil,mulimani2025closer}.
 Suppose speech is learned at the first phase and vehicle sounds at the subsequent phase. A next-phase recording may contain both speech and a passing car, yet only the vehicle classes are annotated. A missing speech label therefore does not indicate absence: the old-class annotation is simply unavailable. Training on only the annotated new classes thus pushes the model to actively suppress old-class predictions~\cite{dong2023krt,zhang2025la}, accelerating catastrophic
forgetting~\cite{mccloskey1989catastrophic,masana2022survey} rather than failing to reinforce it.


Continual learning methods mitigate catastrophic forgetting through
replay~\cite{rebuffi2017icarl}, parameter
regularization, or knowledge distillation~\cite{kirkpatrick2017ewc,
zenke2017continual,li2017lwf}, and recent work has extended these to
multi-label CIL~\cite{dong2023krt,kim2024csc,demin2024multilane}.
However, all rely on iterative gradient updates, which are particularly
problematic in the multi-label ICL regime: 
missing annotations for old
classes are interpreted as indicating absent classes, 
producing suppressive gradients that actively
overwrite old knowledge. Knowledge distillation partially alleviates
this, but the old model has never observed new-class sounds, making its
targets unreliable precisely where the ambiguity is most severe.


Analytic continual learning instead derives the classifier in closed
form via recursive least squares~\cite{zhuang2022acil}, and has been explored for single-label audio tasks
including keyword spotting, sound-source localization, and audio
deepfake detection~\cite{xiao2025analytickws,fan2026analytic, xiao2025listen}. Crucially, missing old-class labels contribute zero
to the closed-form update, leaving old-class weights unchanged by
construction: a property gradient-based methods cannot replicate.
Whether this advantage extends to multi-label audio CIL, where labels
cover only the current class group and leave old sound events
unannotated, remains an open question.

\begin{figure*}[t!]
\centering
\includegraphics[width=\linewidth]{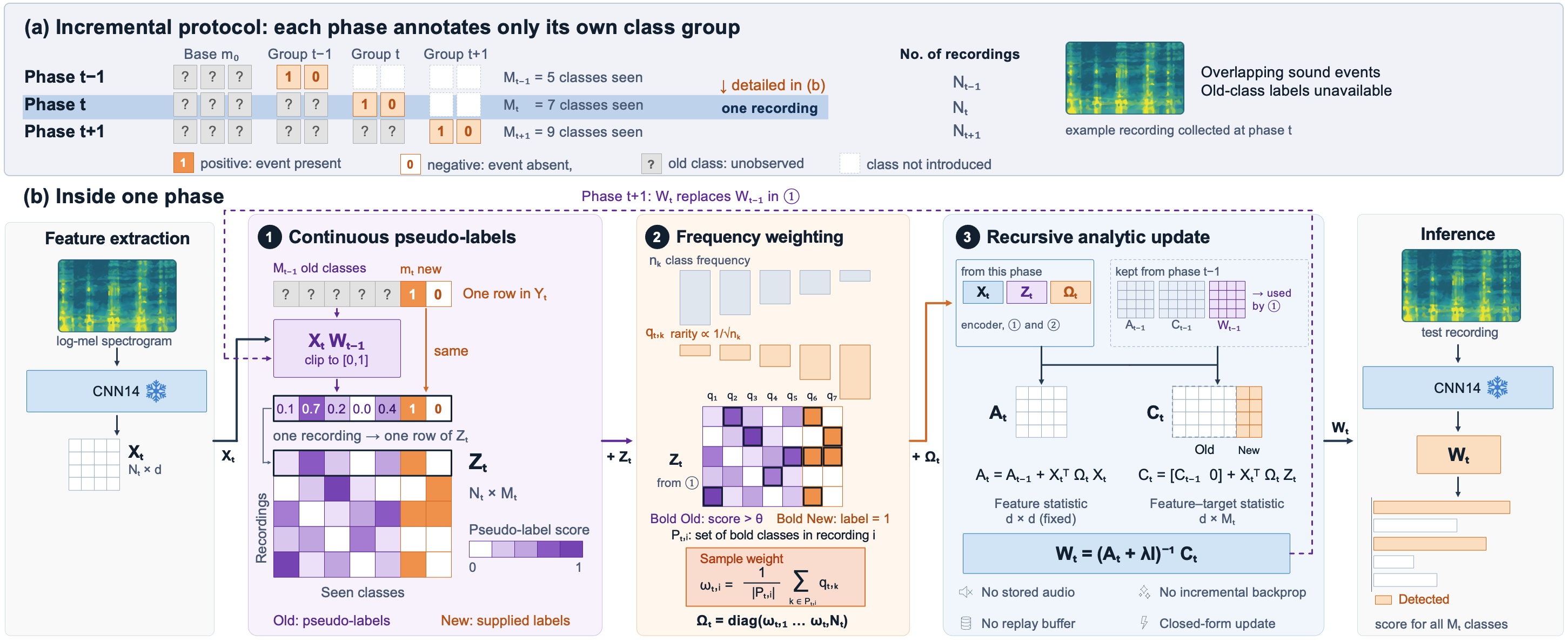}
\caption{Overview of \textsc{\method}. \textbf{(a)} At each phase only the new
class group is annotated; old classes are unobserved,
not negative. \textbf{(b)} Within phase~$t$, a frozen CNN14 gives $\mathbf{X}_t$;
(1) clipped scores $\mathbf{X}_t\mathbf{W}_{t-1}$ and $\mathbf{Y}_t$ complete the
target matrix $\mathbf{Z}_t$; (2) rarity-based weights form
$\boldsymbol{\Omega}_t$; (3) $\mathbf{X}_t,\mathbf{Z}_t,\boldsymbol{\Omega}_t$
update $\mathbf{A}_t,\mathbf{C}_t$ recursively, giving $\mathbf{W}_t$ in closed
form and returning it to step~(1) at the next phase. Inference scores all $M_t$
classes with the frozen encoder and $\mathbf{W}_t$.}
\label{fig:overview}
\vspace{-4mm}
\end{figure*}

To investigate this question, we propose an analytic framework for multi-label audio CIL that exploits the key property of the closed-form update leaving old-class weights intact by construction. 
Building on this, we further propose \method (\textbf{A}nalytic \textbf{L}earning for \textbf{M}ulti-label \textbf{A}udio), which extends the framework~\cite{zhang2025la} with two mechanisms tailored to the multi-label regime. Rather than accepting zero-filled old-class targets, the previous-phase classifier generates soft predictions (i.e., pseudo-label) for old classes on current-phase recordings, providing a more faithful estimate of which old sounds are actually present. 
Frequency-based sample weighting is additionally incorporated to prevent dominant sound classes from skewing the classifier update~\cite{cui2019classbalanced}. To our knowledge, this is the first work to explore analytic continual learning for multi-label audio CIL. We also provide systematic evaluation to validate its effectiveness.


The findings show that analytic learning yields consistent improvements over gradient-based baselines, and \method with soft pseudo-labels and frequency weighting further demonstrates the benefit of explicitly addressing unannotated old classes. This suggests that the closed-form update is inherently better suited to the multi-label ICL regime than iterative gradient optimization, and that analytic continual learning is a promising direction for multi-label audio CIL in real-world deployments where exhaustive annotation across all sound classes is impractical\footnote{The code will be released publicly.}.


\section{\method for Audio Multi-Label CIL}
\label{sec:method}
\subsection{Problem Setting and Notation}
\label{sec:problem_setting}
We follow the established multi-label class-incremental learning
protocol~\cite{mulimani2024mlaudiocil,mulimani2025closer,zhang2025la},
in which classes are introduced over a sequence of phases. At phase $t$,
training annotations are provided only for the class group introduced in
that phase. However, because an audio recording can contain several
overlapping sound events, a current recording may also contain events from
previously learned classes. Their annotations are unavailable during the
current phase. 


Formally, as shown in Figure~\ref{fig:overview}(a), the learning process unfolds over phases $t = 0, 1, \ldots, T$.
At each phase~$t$, a new group of $m_t$ sound classes is introduced,
and the learner has seen $M_t = \sum_{i=0}^{t} m_i$ classes in total by
the end of that phase. At phase~$t$, the learner receives $N_t$ audio recordings. Each recording
is passed through a frozen audio encoder to produce a $d$-dimensional
feature vector, and the full set of recordings is represented as a feature
matrix $\mathbf{X}_t \in \mathbb{R}^{N_t \times d}$ where each row corresponds to one recording.

The annotations at phase~$t$ cover only the $m_t$ newly
introduced classes, forming a binary label matrix $\mathbf{Y}_t \in \{0,1\}^{N_t \times m_t}$ where 1 denotes a positive label indicating the presence of a newly
introduced sound class, and 0 denotes a negative label indicating its absence.
$\mathbf{Y}_t$ has no columns for the $M_{t-1}$ classes learned in previous phases. 

To train a classifier over all $M_t$ classes seen so far, we construct a
completed target matrix $\mathbf{Z}_t \in [0,1]^{N_t \times M_t}$ whose columns are assembled from two sources: the current annotations
$\mathbf{Y}_t$ fill the columns for new classes, while the classifier
from phase~$t-1$ supplies soft score estimates for the old-class columns.
The analytic classifier $\mathbf{W}_t \in \mathbb{R}^{d \times M_t}$ is then fitted to map $\mathbf{X}_t$ to $\mathbf{Z}_t$ in closed form,
and produces a score for each of the $M_t$ classes for any input.




\subsection{Analytic Learning}
\label{sec:preliminaries}

As shown in Figure~\ref{fig:overview}(b) (left), given $N_t$ audio recordings at phase~$t$, the encoder $f_\theta$
extracts a $d$-dimensional feature vector $x_i = f_\theta(a_i)$ for
each recording $a_i$, stacked into a feature matrix
$\mathbf{X}_t \in \mathbb{R}^{N_t \times d}$. A linear classifier
$\mathbf{W}_t \in \mathbb{R}^{d \times M_t}$ maps these features to
scores over all $M_t$ sound classes seen so far. After the base phase,
the encoder is frozen to preserve the learned acoustic representations
across phases, while the linear classifier is updated analytically
using $\mathbf{X}_t$ and the completed target matrix $\mathbf{Z}_t$
to incorporate newly introduced sound classes without gradient updates
or access to historical recordings. Here, the old classes in \(\mathbf{Z}_t\) not annotated are represented as 0.


The classifier $\mathbf{W}_t$ is estimated by minimizing the ridge
regression objective
\begin{equation}
    \min_{\mathbf{W}_t} \ \|\mathbf{X}_t \mathbf{W}_t - \mathbf{Z}_t\|_F^2
    + \lambda \|\mathbf{W}_t\|_F^2,
    \label{eq:basic_objective}
\end{equation}
where the first term minimizes prediction error over all $M_t$ classes
and $\lambda > 0$ controls the regularization strength. This objective
admits the closed-form solution
\begin{equation}
    \mathbf{W}_t = (\mathbf{X}_t^\top \mathbf{X}_t + \lambda I_d)^{-1}
    \mathbf{X}_t^\top \mathbf{Z}_t.
    \label{eq:basic_ridge}
\end{equation}

The solution depends on the training data only through two sufficient
statistics
\begin{equation}
    \mathbf{A}_t = \mathbf{X}_t^\top \mathbf{X}_t \in \mathbb{R}^{d \times d},
    \qquad
    \mathbf{C}_t = \mathbf{X}_t^\top \mathbf{Z}_t \in \mathbb{R}^{d \times M_t},
    \label{eq:basic_statistics}
\end{equation}
so that $\mathbf{W}_t = (\mathbf{A}_t + \lambda I_d)^{-1} \mathbf{C}_t$.
When phase~$t$ arrives, these statistics are accumulated as
\begin{equation}
    \mathbf{A}_t \leftarrow \mathbf{A}_{t-1} + \mathbf{X}_t^\top \mathbf{X}_t,
    \qquad
    \mathbf{C}_t \leftarrow \mathbf{C}_{t-1} + \mathbf{X}_t^\top \mathbf{Z}_t,
    \label{eq:accumulate}
\end{equation}
and $\mathbf{W}_t$ is recomputed from the updated statistics via
Eq.~\eqref{eq:basic_ridge}. This recursive update is mathematically
equivalent to retraining jointly on all phases seen so far, yet requires
neither historical recordings nor back-propagation~\cite{zhuang2022acil}.



\subsection{Proposed \method }
\label{sec:method_overview}

Figure~\ref{fig:overview}(b) presents \textsc{\method}. Building on
the analytic learning framework described above, \textsc{\method}
incorporates two mechanisms to address the missing-label regime of
multi-label audio CIL: continuous pseudo-labeling and frequency-based sample weighting.


\subsubsection{Continuous Pseudo-Labels}
\label{sec:target_construction}

At phase $t \geq 1$, the supplied label matrix $\mathbf{Y}_t$ covers
only the $m_t$ newly introduced classes, leaving old-class labels
unobserved. Rather than filling these with zeros, we use the preceding
classifier $\mathbf{W}_{t-1}$ to estimate them. Specifically, we
compute old-class regression scores
\begin{equation}
    \mathbf{S}_t^{\mathrm{old}} = \mathbf{X}_t \mathbf{W}_{t-1}
    \in \mathbb{R}^{N_t \times M_{t-1}},
    \label{eq:old_scores}
\end{equation}
and clip each entry to $[0,1]$ to obtain soft pseudo-labels
$\widetilde{\mathbf{Y}}_t^{\mathrm{old}}$. Retaining continuous scores
rather than thresholding them preserves uncertainty about which old
sound classes are present. The completed target matrix is then
\begin{equation}
    \mathbf{Z}_t = \left[\,\widetilde{\mathbf{Y}}_t^{\mathrm{old}}
    \;\Big|\; \mathbf{Y}_t\,\right] \in [0,1]^{N_t \times M_t},
    \label{eq:pl_targets}
\end{equation}
where each row contains soft targets for old classes followed by
binary targets for the current class group. At the base phase,
$\mathbf{Z}_0 = \mathbf{Y}_0$.

\subsubsection{Frequency-Based Sample Weighting}
\label{sec:sample_weighting}

Audio events are generally imbalanced, with some events occurring more often and others occurring less frequently. To prevent
dominant classes from skewing the classifier update, each recording is
assigned a weight inversely proportional to the frequency of its
positive classes.

Concretely, each recording $i$ is assigned a scalar weight
$\omega_{t,i}$ that reflects the rarity of its positive sound classes.
Let $n_k$ denote the number of positive recordings for class $k$ at
its introduction phase. The rarity factor of class $k$ is defined as
\begin{equation}
    q_{t,k} = M_t n_k^{-1/2}\,\bigg/\!\sum_{\ell=1}^{M_t} n_\ell^{-1/2},
    \label{eq:class_weights}
\end{equation}
which assigns larger values to less frequent classes, and then normalizes
the mean rarity factor across all $M_t$ classes to
one~\cite{cui2019classbalanced,zhang2025la}. The weight of recording
$i$ is the average rarity factor over its positive classes $P_{t,i}$,
which includes both supplied new-class labels and pseudo-labeled old
classes whose scores exceed a threshold $\theta$:
\begin{equation}
    \omega_{t,i} = \frac{1}{|P_{t,i}|}
    \sum_{k \in P_{t,i}} q_{t,k}.
    \label{eq:recording_weight}
\end{equation}
These per-recording weights are collected into a diagonal matrix, \(\boldsymbol{\Omega}_t = \operatorname{diag}(\omega_{t,1}, \ldots, \omega_{t,N_t})\), and used in the closed-form solution.

\subsubsection{Recursive Analytic Update}
\label{sec:recursive_statistics}
The completed targets $\mathbf{Z}_t$ and sample weights
$\boldsymbol{\Omega}_t$ together define the weighted ridge regression
objective accumulated over all phases:
\begin{equation}
    \mathcal{J}_t(\mathbf{W})
    = \sum_{\tau=0}^{t}
    \left\|\boldsymbol{\Omega}_\tau^{1/2}
    \left(\mathbf{X}_\tau \mathbf{W} - \mathbf{Z}_{\tau\rightarrow t}
    \right)\right\|_F^2
    + \lambda\|\mathbf{W}\|_F^2,
    \label{eq:weighted_ridge}
\end{equation}
where $\mathbf{Z}_{\tau\rightarrow t}$ pads the targets from phase
$\tau$ with zero columns for classes introduced in later phases.
Rather than storing the feature and target matrices from every phase,
the solution to Eq.~\eqref{eq:weighted_ridge} can be recovered from
two sufficient statistics. At the base phase, these are now initialized as
\begin{equation}
    \mathbf{A}_0 = \mathbf{X}_0^\top \boldsymbol{\Omega}_0 \mathbf{X}_0,
    \qquad
    \mathbf{C}_0 = \mathbf{X}_0^\top \boldsymbol{\Omega}_0 \mathbf{Z}_0.
\end{equation}
At each subsequent phase, they are updated as
\begin{equation}
    \mathbf{A}_t = \mathbf{A}_{t-1}
    + \mathbf{X}_t^\top \boldsymbol{\Omega}_t \mathbf{X}_t,
    \label{eq:recursive_A}
\end{equation}
\begin{equation}
\mathbf{C}_t =
\left[\mathbf{C}_{t-1}\;\; \mathbf{0}_{d\times m_t}\right]
+ \mathbf{X}_t^\top \boldsymbol{\Omega}_t \mathbf{Z}_t,
\label{eq:recursive_C}
\end{equation}
where$\left[\mathbf{C}_{t-1}\;\;\mathbf{0}_{d\times m_t}\right]$
denotes the horizontal concatenation of $\mathbf{C}_{t-1}$ and a
zero matrix containing one column for each newly introduced class. 
The retained $\mathbf{C}_{t-1}$
preserves all prior supervision for old classes, while the current
term adds contributions from phase-$t$ recordings. The classifier is
then recovered in closed form as
\begin{equation}
    \mathbf{W}_t = (\mathbf{A}_t + \lambda I_d)^{-1} \mathbf{C}_t.
    \label{eq:analytic_solution}
\end{equation}
Here, the weight is introduced and used to update the solution.

\section{Evaluation Protocol}
\subsection{Data and baselines}
We use a 50-class subset of AudioSet-R~\cite{sun2025audiosetr}, selecting
the 50 most frequent classes ordered from most to least frequent. Each
ten-second clip may carry multiple positive labels and can appear in more
than one phase set. Training at each phase exposes only the current class
group's labels, while evaluation uses all labels for classes seen so far.


Audio is represented by 64-bin log-mel spectrograms. We train
CNN14~\cite{kong2020panns} during the base phase and freeze the encoder
thereafter. Its 4096-dimensional embeddings provide the fixed features.


Baselines include fine-tuning (FT), elastic weight consolidation (EWC),
synaptic intelligence (SI), learning without forgetting (LwF), Joint
training, and per-phase retraining (PPR)~\cite{kirkpatrick2017ewc,zenke2017continual,li2017lwf}.
We adapt these gradient-based methods to multi-label audio using independent
per-class sigmoid outputs and binary cross-entropy. FT uses only current-phase clips and serves as a lower-bound reference. Joint and PPR access historical recordings, while EWC, SI, and LwF are regularization and distillation baselines. Hyperparameters are selected
on a separate validation set. Average precision (AP) measures the ranking
of positive clips for one class, and mean average precision (mAP) averages
AP across classes. Cumulative mAP covers all classes observed at each
phase. Mean cum. averages these phase scores. Final mAP is the last-phase
score.

\section{Results}
\subsection{Overall Incremental Performance}
Table~\ref{tab:main} summarizes the results. Analytic CIL, which uses
zero-filled old-class targets without $\boldsymbol{\Omega}$, already
achieves 42.73\% mean cumulative mAP and 38.43\% final mAP, outperforming
all gradient-based methods. The strongest of these, LwF,
obtains 37.95\% and 26.98\%, respectively. \method further raises these
by 0.59 and 0.65 percentage points to 43.32\% and 39.08\%. The much
larger gap between gradient-based methods and Analytic CIL shows that the
analytic framework shows benefits in this task, and the
pseudo-labeling and frequency weighting components provide additional gain.

\begin{figure}[t]
\centering
\includegraphics[width=0.6\linewidth]{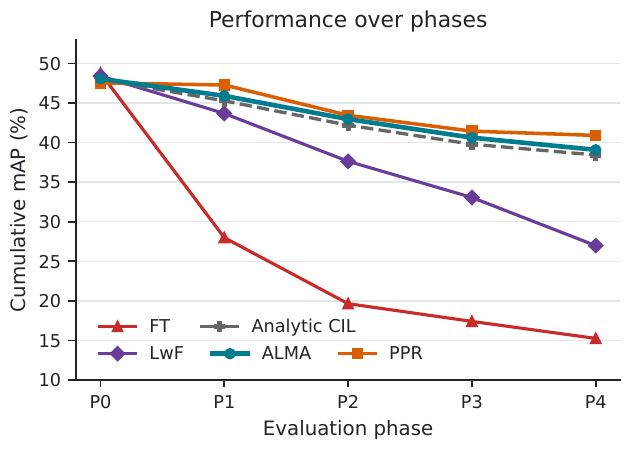}
\vspace{-2mm}
\caption{Cumulative mAP across evaluation phases. The comparison shows
how the advantage of the analytic methods emerges as new class groups
are introduced.}
\label{fig:main}
\end{figure}

The strong performance of the analytic methods can be attributed to how
they incorporate knowledge from previous phases. Gradient-based methods
optimize only on current-phase data at each update, which risks
overwriting what was learned before. The analytic framework instead
accumulates all previous phases exactly into a single closed-form
solution, so no previously learned information is discarded. This is
further supported by the comparison with PPR, which retrains on all
historical data at every phase and serves as an upper bound. \method
closes most of this gap, scoring only 0.80 and 1.82 points below PPR
despite storing no historical recordings. As shown in
Figure~\ref{fig:main}, \method starts below FT and LwF at the base
phase but surpasses both after the first incremental update, confirming
that its advantage grows as more phases are observed.


\begin{table}[t]
\centering
\caption{
Performance comparison with baselines.}
\label{tab:main}
\small
\resizebox{0.9\columnwidth}{!}{
\begin{tabular}{lccrr}
\toprule
Method & Past data & Inc. BP & Mean mAP & Final mAP \\
\midrule
FT & \xmark & \cmark & 25.82 & 15.26 \\
EWC~\cite{kirkpatrick2017ewc} & \xmark & \cmark & 25.91 & 14.96 \\
SI~\cite{zenke2017continual} & \xmark & \cmark & 33.22 & 25.25 \\
LwF~\cite{li2017lwf} & \xmark & \cmark & 37.95 & 26.98 \\
\midrule
\textbf{Analytic CIL (Ours)} & \xmark & \xmark & 42.73 & 38.43 \\
\textbf{\method (Ours)} & \xmark & \xmark & \textbf{43.32} & \textbf{39.08} \\
\midrule
Joint & \cmark & -- & 43.46 & 40.45 \\
PPR & \cmark & \cmark & 44.12 & 40.90 \\
\bottomrule
\end{tabular}
}
\end{table}

\setcounter{figure}{2}
\begin{figure}[t!]
\begin{minipage}{\columnwidth}
\centering
\includegraphics[width=0.55\linewidth]{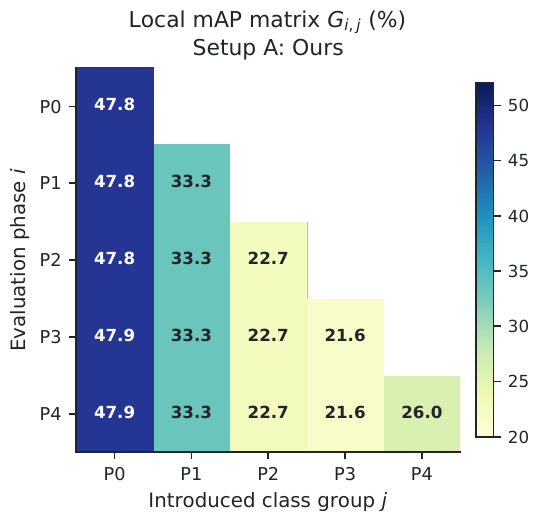}
\vspace{-2mm}
\captionof{figure}{Local-mAP matrix $G_{i,j}$ on a fixed test pool. Row $i$
is the evaluation phase and column $j$ the class group; blanks are groups not
yet introduced.}
\label{fig:cl}
\end{minipage}
\end{figure}

\setcounter{figure}{3}
\begin{figure}[t]
\centering
\includegraphics[width=0.7\linewidth]{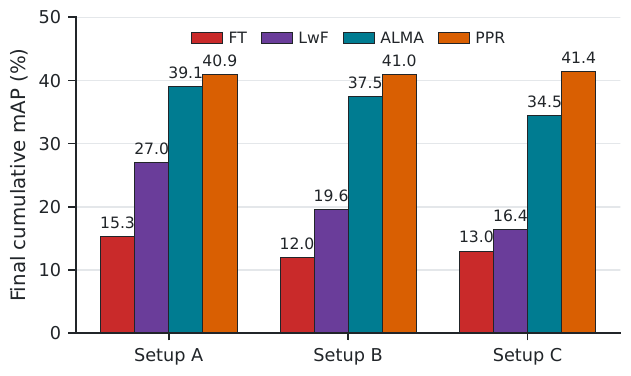}
\vspace{-2mm}
\caption{Final cumulative mAP (\%) across Setups A--C with increasingly
long incremental sequences.}
\label{fig:robust}
\end{figure}

\subsection{New-Class Learning and Old-Class Retention}
Figure~\ref{fig:cl} presents the local mAP of each class group across
phases, where each column tracks one group from its introduction
onward. If the model forgets previously learned classes as new ones
are added, local mAP would decline across a column. Instead, the base
group (P0) maintains approximately 47.8\% local mAP through all
subsequent phases, and the group introduced at P1 similarly holds at
approximately 33.3\%. This stability across all class groups
demonstrates that \method retains previously learned classes as new
ones added, confirming strong resistance to catastrophic forgetting.



\subsection{Performance Across Incremental Protocols}
\label{sec:schedule_results}
Figure~\ref{fig:robust} evaluates whether the advantage of \method
persists as the base phase shrinks and the number of incremental phases
grows. Setup~A begins with 30 base classes followed by four five-class
phases; Setup~B with 20 followed by six; and Setup~C with 10 followed
by eight: providing a progressively more challenging evaluation.
\method achieves final mAP of 39.08\%, 37.47\%, and 34.50\% across
Setups~A--C, consistently outperforming LwF by 12.10, 17.89, and 18.13
points and FT by 23.82, 25.49, and 21.47 points. It also remains close
to PPR, the historical-data upper bound, with gaps of only 1.82, 3.54,
and 6.93 points. The widening gap to PPR reflects the increasing
difficulty of later setups, but the advantage over gradient-based
methods remains consistent throughout.


\subsection{Ablation Study}
\label{sec:target_weight_results}
Table~\ref{tab:omega_ablation} presents a matched ablation isolating
the contributions of old-class target construction and sample weighting
beyond the basic analytic learner. Hard PL denotes binarizing
the preceding classifier's old-class scores at a threshold of 0.5,
whereas Continuous PL retains the clipped scores in $[0,1]$. Replacing zero-filled old-class
targets with hard pseudo-labels improves mean cumulative and final mAP
by 0.29 and 0.28 points, and retaining continuous scores instead of
hard labels adds a further 0.27 and 0.40 points.
Adding sample
weighting $\boldsymbol{\Omega}$ to continuous targets changes
mean and final mAP by $+0.03$ and $-0.03$ points respectively. These results
show that old-class target construction provides the main additional
gain beyond the basic analytic learner, with a modest advantage from
continuous over hard pseudo-labels, while sample weighting provides no
consistent additional benefit.

\begin{table}[t]
\centering
\caption{Setup A ablation. PL denotes pseudo-labels; Mean averages P0--P4.}
\label{tab:omega_ablation}
\scriptsize
\setlength{\tabcolsep}{1.5pt}
\renewcommand{\arraystretch}{0.85}
\resizebox{\columnwidth}{!}{\begin{tabular}{@{}llc*{6}{r}@{}}
\toprule
Variant & Old-class target & $\Omega$ & P0 & P1 & P2 & P3 & P4 & Mean \\
\midrule
Analytic CIL
& Zero-filled & \xmark
& 47.968 & 45.270 & 42.203 & 39.791 & 38.429 & 42.732 \\
\midrule
\multirow{4}{*}{Ablations}
& Zero-filled & \cmark
& 48.045 & 44.702 & 41.528 & 39.086 & 37.862 & 42.245 \\
& Hard PL & \xmark
& 47.968 & 45.639 & 42.638 & 40.178 & 38.706 & 43.026 \\
& Hard PL & \cmark
& 48.045 & 45.705 & 42.659 & 40.177 & 38.700 & 43.057 \\
& Continuous PL & \xmark
& 47.968 & 45.829 & 42.943 & \textbf{40.622}
& \textbf{39.111} & 43.295 \\
\midrule
ALMA
& Continuous PL & \cmark
& \textbf{48.045} & \textbf{45.905} & \textbf{42.962}
& 40.619 & 39.084 & \textbf{43.323} \\
\bottomrule
\end{tabular}
}
\end{table}


\section{Conclusion}
We investigated analytic continual learning for multi-label audio CIL
without historical recordings or incremental back-propagation. The
analytic learner substantially outperforms gradient-based methods across
all evaluated setups, identifying the closed-form recursive update as
the main source of improvement. \method further improves performance
by replacing zero-filled targets with continuous old-class score
estimates, demonstrating that better target construction consistently
benefits the analytic framework. Local mAP of previously learned class
groups remains nearly unchanged as new classes are added, confirming
strong resistance to catastrophic forgetting. These results establish
analytic learning as an effective and computationally simple direction
for multi-label audio CIL.


\bibliographystyle{IEEEbib}
\bibliography{references}

@inproceedings{dekkers2017sins,
  title     = {The {SINS} Database for Detection of Daily Activities in a Home Environment Using an Acoustic Sensor Network},
  author    = {Dekkers, Gert and Lauwereins, Steven and Thoen, Bart and Adhana, Mulu Weldegebreal and Brouckxon, Henk and van Waterschoot, Toon and Vanrumste, Bart and Verhelst, Marian and Karsmakers, Peter},
  booktitle = {Proc. DCASE},
  pages     = {32--36},
  year      = {2017},
  url       = {https://dcase.community/documents/workshop2017/proceedings/DCASE2017Workshop_Dekkers_32.pdf}
}

@inproceedings{purohit2019mimii,
  title     = {{MIMII} Dataset: Sound Dataset for Malfunctioning Industrial Machine Investigation and Inspection},
  author    = {Purohit, Harsh and Tanabe, Ryo and Ichige, Takeshi and Endo, Takashi and Nikaido, Yuki and Suefusa, Kaori and Kawaguchi, Yohei},
  booktitle = {Proc. DCASE},
  pages     = {209--213},
  year      = {2019},
  doi       = {10.33682/m76f-d618},
  url       = {https://doi.org/10.33682/m76f-d618}
}

@article{kahl2021birdnet,
  title   = {{BirdNET}: A Deep Learning Solution for Avian Diversity Monitoring},
  author  = {Kahl, Stefan and Wood, Connor M. and Eibl, Maximilian and Klinck, Holger},
  journal = {Ecological Informatics},
  volume  = {61},
  pages   = {101236},
  year    = {2021},
  doi     = {10.1016/j.ecoinf.2021.101236},
  url     = {https://doi.org/10.1016/j.ecoinf.2021.101236}
}

@article{mohmmad2024sedreview,
  author  = {Sallauddin Mohmmad and Suresh Kumar Sanampudi},
  title   = {Exploring Current Research Trends in Sound Event Detection: A Systematic Literature Review},
  journal = {Multimedia Tools and Applications},
  volume  = {83},
  number  = {37},
  pages   = {84699--84741},
  year    = {2024},
  doi     = {10.1007/s11042-024-18740-9},
  url     = {https://doi.org/10.1007/s11042-024-18740-9}
}

@inproceedings{joshi2018timeaggregation,
  author    = {Joshi, Pankaj and Gautam, Digvijaysingh and Ramakrishnan, Ganesh and Jyothi, Preethi},
  title     = {Time Aggregation Operators for Multi-label Audio Event Detection},
  booktitle = {Proc. Interspeech},
  pages     = {3309--3313},
  year      = {2018},
  doi       = {10.21437/Interspeech.2018-1637},
  url       = {https://doi.org/10.21437/Interspeech.2018-1637}
}

@inproceedings{demin2024multilane,
  title     = {Less is more: Summarizing Patch Tokens for efficient Multi-Label Class-Incremental Learning},
  author    = {De Min, Thomas and Mancini, Massimiliano and Lathuili{\`e}re, St{\'e}phane and Roy, Subhankar and Ricci, Elisa},
  booktitle = {Proceedings of the 3rd Conference on Lifelong Learning Agents},
  series    = {Proceedings of Machine Learning Research},
  volume    = {274},
  pages     = {852--868},
  publisher = {PMLR},
  year      = {2025},
  url       = {https://proceedings.mlr.press/v274/min25a.html}
}

@inproceedings{kim2024csc,
  title     = {Confidence Self-Calibration for Multi-Label Class-Incremental Learning},
  author    = {Du, Kaile and Zhou, Yifan and Lyu, Fan and Li, Yuyang and Lu, Chen and Liu, Guangcan},
  booktitle = {Proc. ECCV},
  pages     = {234--252},
  year      = {2024},
  doi       = {10.1007/978-3-031-72751-1_14},
  url       = {https://doi.org/10.1007/978-3-031-72751-1_14}
}

@inproceedings{dong2023krt,
  title     = {Knowledge Restore and Transfer for Multi-Label Class-Incremental Learning},
  author    = {Dong, Songlin and Luo, Haoyu and He, Yuhang and Wei, Xing and Cheng, Jie and Gong, Yihong},
  booktitle = {Proc. IEEE/CVF ICCV},
  pages     = {18711--18720},
  year      = {2023},
  doi       = {10.1109/ICCV51070.2023.01715},
  url       = {https://openaccess.thecvf.com/content/ICCV2023/html/Dong_Knowledge_Restore_and_Transfer_for_Multi-Label_Class-Incremental_Learning_ICCV_2023_paper.html}
}

@inproceedings{fan2026analytic,
  title     = {Analytic Incremental Learning for Sound Source Localization with Imbalance Rectification},
  author    = {Fan, Zexia and Chen, Yu and Zhang, Qiquan and Chen, Kainan and Qian, Xinyuan},
  booktitle = {Proc. IEEE ICASSP},
  pages     = {20991--20995},
  year      = {2026},
  doi       = {10.1109/ICASSP55912.2026.11461645},
  url       = {https://doi.org/10.1109/ICASSP55912.2026.11461645}
}

@inproceedings{xiao2025analytickws,
  title     = {{AnalyticKWS}: Towards Exemplar-Free Analytic Class Incremental Learning for Small-footprint Keyword Spotting},
  author    = {Xiao, Yang and Tianyi, Peng and Das, Rohan Kumar and Hu, Yuchen and Zhuang, Huiping},
  booktitle = {Findings of ACL},
  pages     = {14147--14158},
  year      = {2025},
  publisher = {Association for Computational Linguistics},
  doi       = {10.18653/v1/2025.findings-acl.728},
  url       = {https://aclanthology.org/2025.findings-acl.728/}
}

@inproceedings{gemmeke2017audioset,
  title     = {{Audio Set}: An Ontology and Human-Labeled Dataset for Audio Events},
  author    = {Gemmeke, Jort F. and Ellis, Daniel P. W. and Freedman, Dylan and Jansen, Aren and Lawrence, Wade and Moore, R. Channing and Plakal, Manoj and Ritter, Marvin},
  booktitle = {Proc. IEEE ICASSP},
  pages     = {776--780},
  year      = {2017},
  doi       = {10.1109/ICASSP.2017.7952261},
  url       = {https://doi.org/10.1109/ICASSP.2017.7952261}
}

@article{kirkpatrick2017ewc,
  title   = {Overcoming catastrophic forgetting in neural networks},
  author  = {Kirkpatrick, James and Pascanu, Razvan and Rabinowitz, Neil C. and Veness, Joel and Desjardins, Guillaume and Rusu, Andrei A. and Milan, Kieran and Quan, John and Ramalho, Tiago and Grabska-Barwi{\'n}ska, Agnieszka and Hassabis, Demis and Clopath, Claudia and Kumaran, Dharshan and Hadsell, Raia},
  journal = {Proceedings of the National Academy of Sciences},
  volume  = {114},
  number  = {13},
  pages   = {3521--3526},
  year    = {2017},
  doi     = {10.1073/pnas.1611835114}
}

@article{li2017lwf,
  title   = {Learning without Forgetting},
  author  = {Li, Zhizhong and Hoiem, Derek},
  journal = {IEEE Transactions on Pattern Analysis and Machine Intelligence},
  volume  = {40},
  number  = {12},
  pages   = {2935--2947},
  year    = {2018},
  doi     = {10.1109/TPAMI.2017.2773081}
}

@article{masana2022survey,
  title   = {Class-Incremental Learning: Survey and Performance Evaluation on Image Classification},
  author  = {Masana, Marc and Liu, Xialei and Twardowski, Bart{\l}omiej and Menta, Mikel and Bagdanov, Andrew D. and van de Weijer, Joost},
  journal = {IEEE Transactions on Pattern Analysis and Machine Intelligence},
  volume  = {45},
  number  = {5},
  pages   = {5513--5533},
  year    = {2023},
  doi     = {10.1109/TPAMI.2022.3213473},
  url     = {https://doi.org/10.1109/TPAMI.2022.3213473}
}

@incollection{mccloskey1989catastrophic,
  title     = {Catastrophic Interference in Connectionist Networks: The Sequential Learning Problem},
  author    = {McCloskey, Michael and Cohen, Neal J.},
  booktitle = {Psychology of Learning and Motivation},
  volume    = {24},
  pages     = {109--165},
  publisher = {Academic Press},
  year      = {1989},
  doi       = {10.1016/S0079-7421(08)60536-8},
  url       = {https://doi.org/10.1016/S0079-7421(08)60536-8}
}

@article{kong2020panns,
  title   = {{PANNs}: Large-Scale Pretrained Audio Neural Networks for Audio Pattern Recognition},
  author  = {Kong, Qiuqiang and Cao, Yin and Iqbal, Turab and Wang, Yuxuan and Wang, Wenwu and Plumbley, Mark D.},
  journal = {IEEE/ACM Transactions on Audio, Speech, and Language Processing},
  volume  = {28},
  pages   = {2880--2894},
  year    = {2020},
  doi     = {10.1109/TASLP.2020.3030497}
}

@inproceedings{rebuffi2017icarl,
  title     = {{iCaRL}: Incremental Classifier and Representation Learning},
  author    = {Rebuffi, Sylvestre-Alvise and Kolesnikov, Alexander and Sperl, Georg and Lampert, Christoph H.},
  booktitle = {IEEE Conference on Computer Vision and Pattern Recognition (CVPR)},
  pages     = {5533--5542},
  year      = {2017},
  doi       = {10.1109/CVPR.2017.587},
  url       = {https://doi.org/10.1109/CVPR.2017.587}
}

@inproceedings{mulimani2024mlaudiocil,
  title     = {Class-Incremental Learning for Multi-Label Audio Classification},
  author    = {Mulimani, Manjunath and Mesaros, Annamaria},
  booktitle = {Proc. IEEE ICASSP},
  pages     = {916--920},
  year      = {2024},
  doi       = {10.1109/ICASSP48485.2024.10447952},
  url       = {https://doi.org/10.1109/ICASSP48485.2024.10447952}
}

@article{mulimani2025closer,
  title   = {A Closer Look at Class-Incremental Learning for Multi-Label Audio Classification},
  author  = {Mulimani, Manjunath and Mesaros, Annamaria},
  journal = {IEEE Transactions on Audio, Speech and Language Processing},
  volume  = {33},
  pages   = {1293--1306},
  year    = {2025},
  doi     = {10.1109/TASLPRO.2025.3547233},
  url     = {https://doi.org/10.1109/TASLPRO.2025.3547233}
}

@inproceedings{sun2025audiosetr,
  title     = {{AudioSet-R}: A Refined {AudioSet} with Multi-Stage {LLM} Label Reannotation},
  author    = {Sun, Yulin and Xu, Qisheng and Su, Yi and Zhu, Qian and Dou, Yong and Liu, Xinwang and Xu, Kele},
  booktitle = {Proc. ACM MM},
  pages     = {13089--13096},
  year      = {2025},
  doi       = {10.1145/3746027.3758260}
}

@inproceedings{zhuang2022acil,
  title     = {{ACIL}: Analytic Class-Incremental Learning with Absolute Memorization and Privacy Protection},
  author    = {Zhuang, Huiping and Weng, Zhenyu and Wei, Hongxin and Xie, Renchunzi and Toh, Kar-Ann and Lin, Zhiping},
  booktitle = {Advances in Neural Information Processing Systems (NeurIPS)},
  volume    = {35},
  pages     = {11602--11614},
  year      = {2022},
  url       = {https://arxiv.org/abs/2205.14922}
}

@inproceedings{xiao2022continual,
  title     = {Continual Learning for On-Device Environmental Sound Classification},
  author    = {Xiao, Yang and Liu, Xubo and King, James and Singh, Arshdeep and Chng, Eng Siong and Plumbley, Mark D. and Wang, Wenwu},
  booktitle = {Proc. DCASE},
  year      = {2022},
  url       = {https://dcase.community/documents/workshop2022/proceedings/DCASE2022Workshop_Xiao_47.pdf}
}

@inproceedings{xiao2025listen,
  title     = {Listen, Analyze, and Adapt to Learn New Attacks: An Exemplar-Free Class Incremental Learning Method for Audio Deepfake Source Tracing},
  author    = {Xiao, Yang and Das, Rohan Kumar},
  booktitle = {Interspeech 2025},
  pages     = {1563--1567},
  year      = {2025},
  doi       = {10.21437/Interspeech.2025-16},
  url       = {https://www.isca-archive.org/interspeech_2025/xiao25c_interspeech.html}
}

@inproceedings{zenke2017continual,
  title={Continual Learning Through Synaptic Intelligence},
  author={Zenke, Friedemann and Poole, Ben and Ganguli, Surya},
  booktitle={Proc. ICML},
  volume={70},
  pages={3987--3995},
  year={2017},
  publisher={PMLR},
  url={https://proceedings.mlr.press/v70/zenke17a.html}
}

@inproceedings{zhang2025la,
  title     = {{L3A}: Label-Augmented Analytic Adaptation for Multi-Label Class Incremental Learning},
  author    = {Zhang, Xiang and He, Run and Jiao, Chen and Fang, Di and Li, Ming and Zeng, Ziqian and Chen, Cen and Zhuang, Huiping},
  booktitle = {Proc. ICML},
  series    = {Proceedings of Machine Learning Research},
  volume    = {267},
  pages     = {74938--74949},
  publisher = {PMLR},
  year      = {2025},
  url       = {https://proceedings.mlr.press/v267/zhang25y.html}
}

@inproceedings{cui2019classbalanced,
  title     = {Class-Balanced Loss Based on Effective Number of Samples},
  author    = {Cui, Yin and Jia, Menglin and Lin, Tsung-Yi and Song, Yang and Belongie, Serge},
  booktitle = {IEEE/CVF Conference on Computer Vision and Pattern Recognition (CVPR)},
  pages     = {9268--9277},
  year      = {2019},
  doi       = {10.1109/CVPR.2019.00949},
  url       = {https://openaccess.thecvf.com/content_CVPR_2019/html/Cui_Class-Balanced_Loss_Based_on_Effective_Number_of_Samples_CVPR_2019_paper.html}
}
\end{document}